# DYNAMIC PID LOOP CONTROL

L. Pei, A. Klebaner, J. Theilacker, W. Soyars, A. Martinez,
R. Bossert, B. DeGraff, C. Darve

Fermi National Accelerator Laboratory
Batavia, IL, 60510, USA

## ABSTRACT

The Horizontal Test Stand (HTS) SRF Cavity and Cryomodule 1 (CM1) of eight 9-cell, 1.3GHz SRF cavities are operating at Fermilab. For the cryogenic control system, how to hold liquid level constant in the cryostat by regulation of its Joule-Thompson JT-valve is very important after cryostat cool down to 2.0 K.
  The 72-cell cryostat liquid level response generally takes a long time delay after regulating its JT-valve; therefore, typical PID control loop should result in some cryostat parameter oscillations.
  This paper presents a type of PID parameter self-optimal and Time-Delay control method used to reduce cryogenic system parameters' oscillation.

## INTRODUCTION

The Cryomodule 1 (CM1), an eight 9-cell cavity module operating at 1.3GHz, is known as type III+ Cryomodule 1 and is approximately 12 meter long. A longitudinal view of a type III+ Cryomodule 1 is shown in FIGURE 1 and a 9-cell cavity view is shown in FIGURE 2. The Cryomodule 1 feed configuration is shown in FIGURE 3.
  The fast cool down CM1 to 4.5 K from room temperature ~290 K by fully open PVCD valve (Cv 3.0) will take more than 10 hours after filling CM1 to 50% full liquid helium. And regulation the PVJT valve (Cv 0.12) with pressure around 20 psig will be required to hold CM1 liquid level constant after cool down.
  Due to PVJT valve is smaller size to regulate ~224 liters volume and 12 meters long CM1 cryostat, therefore, CM1's PVJT PID control has significant time-delay. If chosen classic PID feedback loop control, it will cause system oscillation. For constant control task, its control model should be one of the adaptation and dynamic learning PID model.

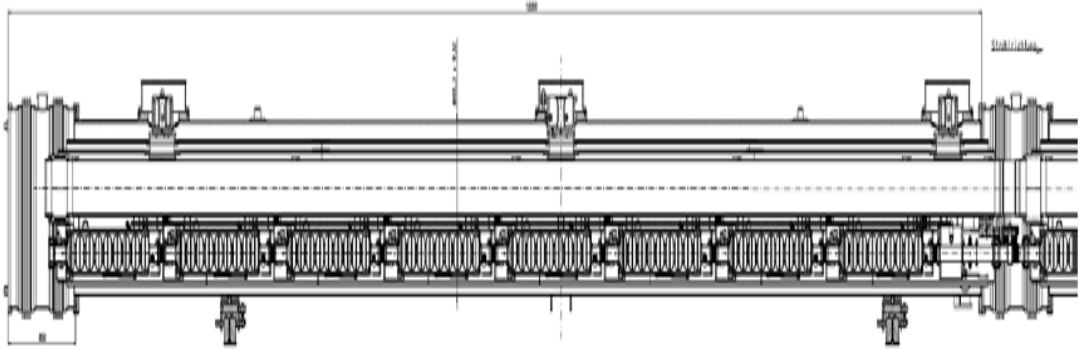

**FIGURE 1.** Longitudinal view of a type III+ Cryomodule 1.

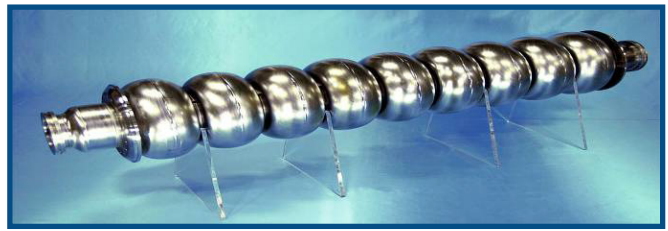

**FIGURE 2.** A 9-cell cavity of a type III+ Cryomodule 1.

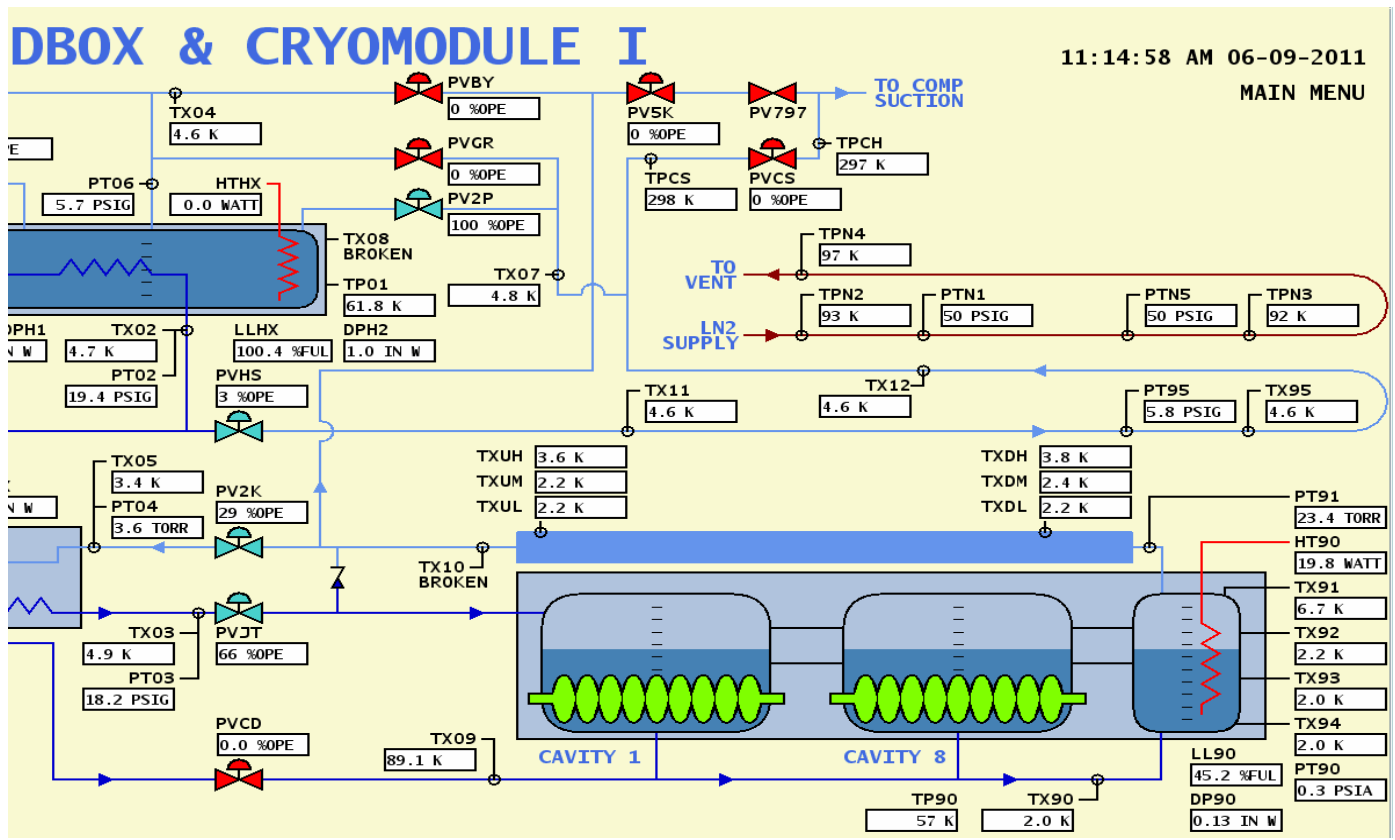

**FIGURE 3.** Cryomodule 1 Feed configuration.

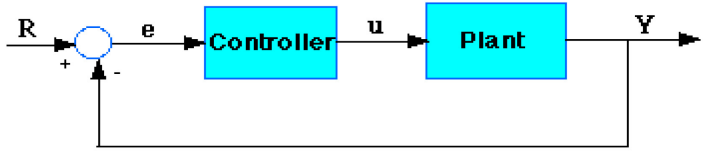

**FIGURE 4.** Classic feedback loop PID control system.

**MATHEMATICAL MODEL**

The classic closed loop PID control system is shown in FIGURE 4. Its transfer function of the PID controller is:

$$u(t) = K_p e(t) + K_I \int e(t)dt + K_D \frac{de(t)}{dt}$$
$$e(t) = R(t) - Y(t)$$

But, for significant time-delay system, time should be broken down to $[-t, -t_2, -t_1, t_0, t]$, lag response time $\Delta t = [-t_1, t_0]$, real-time control PID should be:

Current PID controlled is: $\quad Y(t) = \int_{-t_2}^{-t_1} u(t)dt$

Past PID controlled was: $\quad Y(t) = \int_{-t}^{-t_2} u(t)dt$

Lead or future PID control will be: $\quad Y(t) = \int_{-t_1}^{t_0} u(t)dt$

Therefore, the system transfer function is:

$$Y(t) = \int u(t)dt = \int_{-t}^{-t_2} u(t)dt + \int_{-t_2}^{-t_1} u(t)dt + \int_{-t_1}^{t_0} u(t)dt$$

Real-time learning PID controller transfer function is below.

$$u(-t_2) = K_p(R(-t_2) - Y(-t_2)) + K_I \int_{-t}^{-t_2}(R(-t_2) - Y(-t_2))dt + K_D \frac{d(R(-t_2) - Y(-t_2))}{dt}$$
$$u(-t_1) = K_p(R(-t_1) - Y(-t_1)) + K_I \int_{-t_2}^{-t_1}(R(-t_1) - Y(-t_1))dt + K_D \frac{d(R(-t_1) - Y(-t_1))}{dt}$$
$$u(t_0) = K_p(R(t_0) - Y(t_0)) + K_I \int_{-t_1}^{t_0}(R(t_0) - Y(t_0))dt + K_D \frac{d(R(t_0) - Y(t_0))}{dt}$$

In the real-time learning PID controller, their gains would be change during different period,

$$Y(t) = \int_{-t_1}^{t_0}(K_{P(t_0)}(R(t_0)-Y(t_0)) + K_{I(t_0)}\int_{-t_1}^{t_0}(R(t_0)-Y(t_0))dt + K_{D(t_0)}\frac{d(R(t_0)-Y(t_0))}{dt})dt$$

$$+\int_{-t_2}^{-t_1}(K_{p(-t_1)}(R(-t_1)-Y(-t_1)) + K_{I(-t_1)}\int_{-t_2}^{-t_1}(R(-t_1)-Y(-t_1))dt + K_{D(-t_1)}\frac{d(R(-t_1)-Y(-t_1))}{dt})dt$$

$$+\int_{-t}^{-t_2}(K_{p(-t_2)}(R(-t_2)-Y(-t_2)) + K_{I(-t_2)}\int_{-t}^{-t_2}(R(-t_2)-Y(-t_2))dt + K_{D(-t_2)}\frac{d(R(-t_2)-Y(-t_2))}{dt})dt$$

Since cryogenic cryostat with helium liquid must protect cryogenic system from quench, no big pressure or temperature disturbances are allowed.

If the PID controller with fixed gains controls is based on $t_0$ time only that means it only uses less than 11% of its capability, or loses more than 89% control. Oscillation is foreseeable.

If including $[-t_1 \text{ to } t_0]$, only 33% capability would be used, oscillation is still possible.

If considering $[-t_1 \text{ to } t_0]$ and $[-t_2 \text{ to } -t_1]$ with fixed gains, then the PID control would use less than 66%, so system would be easy to control if it was or has achieved stability already. Then if their gains may adjustable, the 66% capability are used, system would be easier to control.

For ideal control, if analyzing and adjusting all three time periods and their gains between $[-t_1 \text{ to } t_0], [-t_2 \text{ to } -t_1]$ and $[-t \text{ to } -t_2]$, full PID control is applied to the controlled system. The system would be stable, disturbance free and fast to respond, too.

The Dynamic PID loop control system block diagram is shown in FIGURE 5. Its mathematical model is:

$$Y'(t) = K_{t_0}X_{t_0}(t) + K_{-t_1}X_{-t_1}(t) + K_{-t_2}X_{-t_2}(t)$$

And

$$K_{t_{0i}}X_{t_i}(t) = K_{p(t_i)}e(t_i) + K_{I(t_i)}\int_{-t_{i+1}}^{t_i}e(t_i)dt + K_{D(t_i)}\frac{de(t_i)}{dt} \qquad I = 0,-1,-2$$

$K_{t_0}$: Future PID gains, $K_{-t_1}$: Current PID gains, $K_{-t_2}e_{-t_2}(t)$: Constant C for past PID.

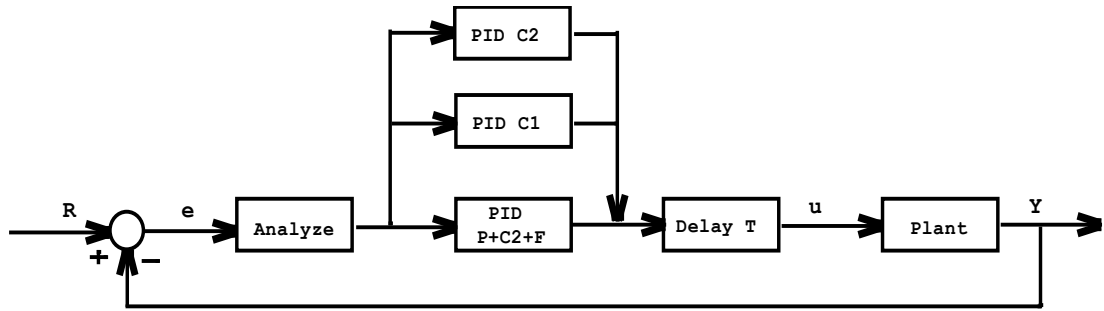

**FIGURE 5.** Dynamic PID loop control system.

# LIQUID LEVEL DYNAMIC PID LOOP CONTROL DESIGN

Siemens APACS real-time, on-line industry control PLC is chosen as our NML CM1 cryogenic control system. Four-Mation software is used as its program tool with four types of languages as Ladder Logic Diagram, Structured Text File, Function Block Diagram and Sequential Function Chart.

In our dynamic and learning PID control model, the helium liquid level L(t), N:3LL90, is chosen as $e_{t_x}(t)$, the JT valve V(t), N:3PVJT, is chosen as $Y(t)$.

$$V'(t) = K_{t_0} L_{t_0}(t) + K_{-t_1} L_{-t_1}(t) + K_{-t_2} L_{-t_2}(t)$$

$$= K_{t_0} L_{t_0}(t) + K_{-t_1} L_{-t_1}(t) + (\sum_{i=-t_2}^{-t} K_i L_i(t_i))/T$$

The future gain $K_{t_0}$ is calculated by exponentiation algorithms $C_i - e^{e(t)}$ of error delta of $(L(t) - SP)$, while delta is less than one. The past PID constant C is calculated by past two hour available time (T). If the error delta is more than one, the current gain $K_{-t_1}$ is working as two responses level to force the PID controller goes to fast responds action.

The CM1 PID control settings are shown in FIGURE 6.

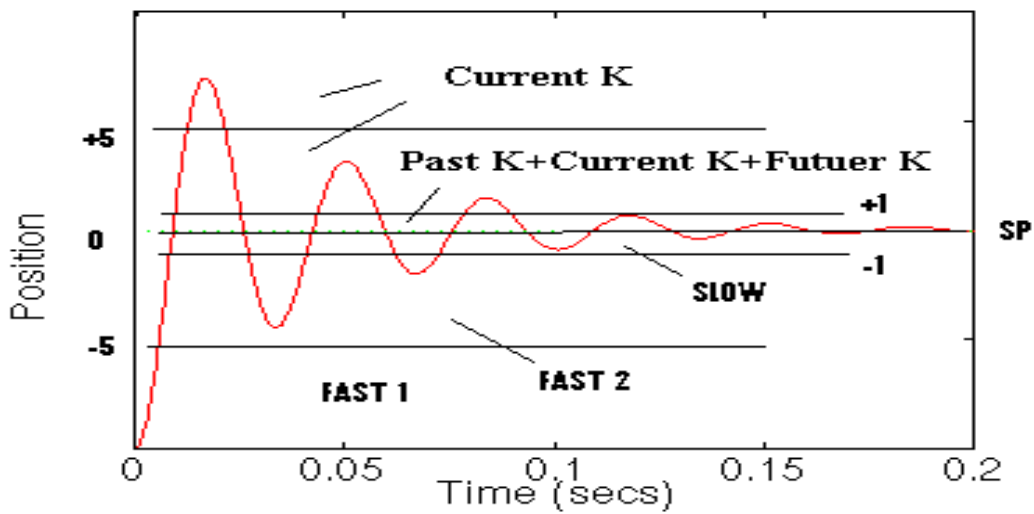

**FIGURE 6.** CM1 PID control settings.

# LIQUID LEVEL DYNAMIC PID LOOP CONTROL TEST AT NML

One example of dynamic PID loop control response for one big RF event is shown in FIGURE 7.

The cryogenic system was in stable status before the big event: helium liquid level, N:3LL90 was at 45%, JT valve, N:3PVJT was at 56% opening, temperature was at 4.5 K, cool-down valve, N:3PVCD was fully closed.

An event suddenly raised N:3LL90 up to 56% more. PID caught it up to close

N:3PVJT Valve to less than 50% opening.

After the event passed, N:3LL90 dropped to 39%. It was over the +/-5% limit setting for 45% set point, so PID starts the first stage faster respond loop as current PID loop control mode.

When N:3LL90 went up to 40% to cross into +/-5% limit for 45% its set point, PID went to second stage the fastest respond loop control. Then valve N:3PVJT opened to its maximum 75% position limit and stayed there.

Once N:3LL90 climbed up to 44%. PID went to slowly response for future, past and current loop control mode. Due to liquid level detector resolution being 0.3%, the PID exponentiation algorithms reduces valve N:3PVJT opening in three steps. If the liquid level detector has better resolution, we will see the valve N:3PVJT opening on continuous decrease exponential curve.

Finally, system helium liquid level N:3LL90 was quickly stable at 45% and valve N:3PVJT stay at 62.5% opening.

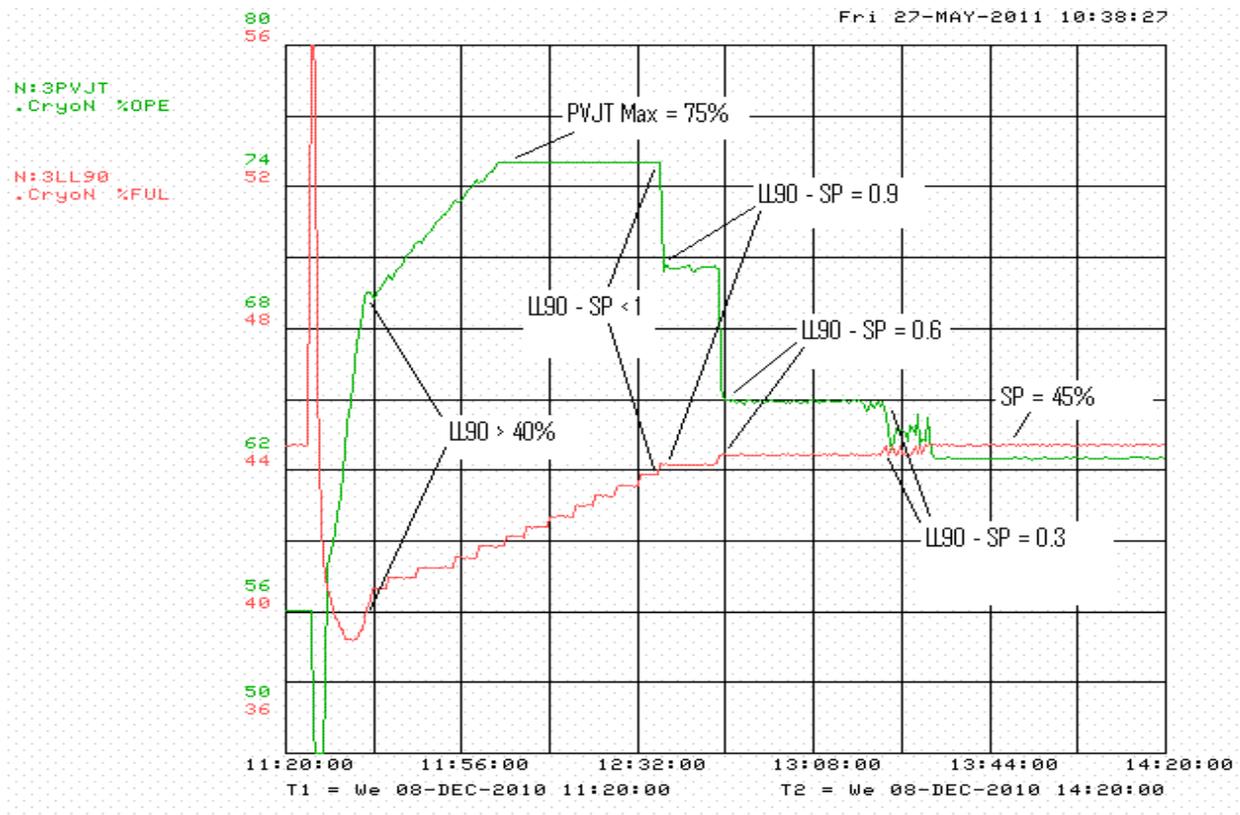

**FIGURE 7.** CM1 PID control response for one big event.

## CONCLUSIONS

The dynamic PID loop control model is especially useful for long-time delay control system. It will dramatically reduce oscillation and quickly stabilize system after some interference happens. How to choose the past, current and future PID gains automatically are very important for the dynamic model. We will continue to study and improve it for many types of control models for Fermilab cryogenic test projects in the future.

## ACKNOWLEDGEMENTS

Fermilab is operated by the Fermi Research Alliance, LLC under Contract No. DE-AC02-07CH11359. The authors wish to recognize the dedication and skills of the Accelerator Cryogenic Department technical personnel involved in the design and testing of the cryogenic equipment.

## REFERENCES

1. Xu, N., Song, W., Xia, A., "System identification", Central South University, Nanjing, China 1991.
2. Norris, B., Bossert, R., Klebaner, A. Lackey, S. Martinez, A., Pei, L., et al., "Cryogenic control for Fermilab's SRF cavitys and test facility", in *Advances in Cryogenics Engineering*, edited by J. G. Weisend, American Institute of Physics, Melville, New York, 2008.
3. Klebaner, A. L., and Theilacker, J. C., "Cryogenics for the Superconducting Module Test Facility," in *Advances in Cryogenics Engineering* 51B, edited by J. G. Weisend, American Institute of Physics, Melville, New York, 2006, pp. 1428-1435.
4. Martinez, A., Klebaner, A.L., Theilacker, J.C., DeGraff, B.D., White, M. and Johnson, G.S., "Fermilab SRF Cryomodule Operational Experience," in *Advances in Cryogenics Engineering*, edited by J. G. Weisend, American Institute of Physics, Melville, New York, 2011, current conference.